\documentclass[journal]{latex}

\usepackage{url}
\usepackage{booktabs}
\usepackage{graphicx,subfigure,array,rotating,amsmath}
\usepackage[table]{xcolor}
\usepackage{cite}

\newcommand {\1}{\columnwidth}
\newcommand {\2}{.48\columnwidth}

\begin{document}
%
\title{Motion model transitions in GPS-IMU sensor fusion for user tracking in augmented reality}
%
%
%

\author{Erkan~Bostanci 
\thanks{E. Bostanci is with Computer Engineering Department, Ankara University, 50. Yil Campus, I Blok, Golbasi, Ankara, 06830, Turkey, e-mail:ebostanci@ankara.edu.tr}
}

%



\maketitle
\thispagestyle{empty}

\begin{abstract}
Finding the position of the user is an important processing step for augmented reality (AR) applications. This paper investigates the use of different motion models in order to choose the  most suitable one, and eventually reduce the Kalman filter errors in sensor fusion for such applications where the accuracy of user tracking is crucial. A Deterministic Finite Automaton (DFA) was employed using the innovation parameters of the filter. Results show that the approach presented here reduces the filter error compared to a static model and prevents filter divergence. The approach was tested on a simple AR game in order to justify the accuracy and performance of the algorithm.
\end{abstract}

\begin{IEEEkeywords}
Deterministic Finite Automaton (DFA), sensor fusion, GPS, IMU, model selection, augmented reality.
\end{IEEEkeywords}

\IEEEpeerreviewmaketitle

\section{Introduction}
\IEEEPARstart{I}{ntegration} of data from Global Positioning System (GPS) and Inertial Measurement Unit (IMU) sensors has been well-studied~\cite{Groves2008, Grigore2014, Barrile2014} in order to improve
upon the robustness of the individual sensors against a number of problems related to accuracy or drift.  The Kalman filter (KF)~\cite{Kalman1960} is the most widely-used filter due to its simplicity and computational efficiency~\cite{Kaplan2005} especially for real-time user tracking applications such as AR. 

Attempts have been made to improve the accuracy of the filter using adaptive values for the state and measurement covariance matrices based on the innovation~\cite{Almagbile2010} and recently fuzzy logic was used for this task~\cite{Tseng2011,Kramer2012}. In some studies~\cite{Ojeda2002,Hong2003} used dynamic motion parameters to decide on the dominance of individual sensors for the final estimate.

Alternative approaches suggest using different motion models for recognizing the type of the motion~\cite{Chen2004,Torr2002,Kanatani2004,Schindler2006,Civera2008b}. Some of these studies (\emph{e.g.}~\cite{Torr2002,Kanatani2004} use a Bayesian framework for identifying a scoring scheme for selecting a motion model and some other studies, see~\cite{Civera2008b}, apply different motion models concurrently and select one of them according to a probabilistic approach.

This paper presents the selection and use of different motion models according to a DFA model~\cite{Bostanci2014} in order to reduce the filter error and ensure faster filter convergence. The rest of the paper is structured as follows: Section~\ref{sec:estimates} presents the methods used for obtaining positional estimates from individual sensors. The fusion filter which uses these motion estimates in order to produce a single output is presented in Section~\ref{sec:filter}. Results are given in Section~\ref{sec:results} and an AR application is presented in Section~\ref{sec:arGame}. Finally, the paper is concluded in Section~\ref{sec:conclusion}.

\section{Finding Position Estimates}
\label{sec:estimates}

Before describing the details of the fusion filter and the DFA approach, it is important to present the calculations used for obtaining individual measurements from the GPS (Phidgets 1040) and IMU sensors (Phidgets Spatial 1056), both low-cost sensors with reasonable accuracy.

\subsection{GPS position estimate}

\begin{figure}[h!t]
  \begin{center}
    \includegraphics[width=\1]{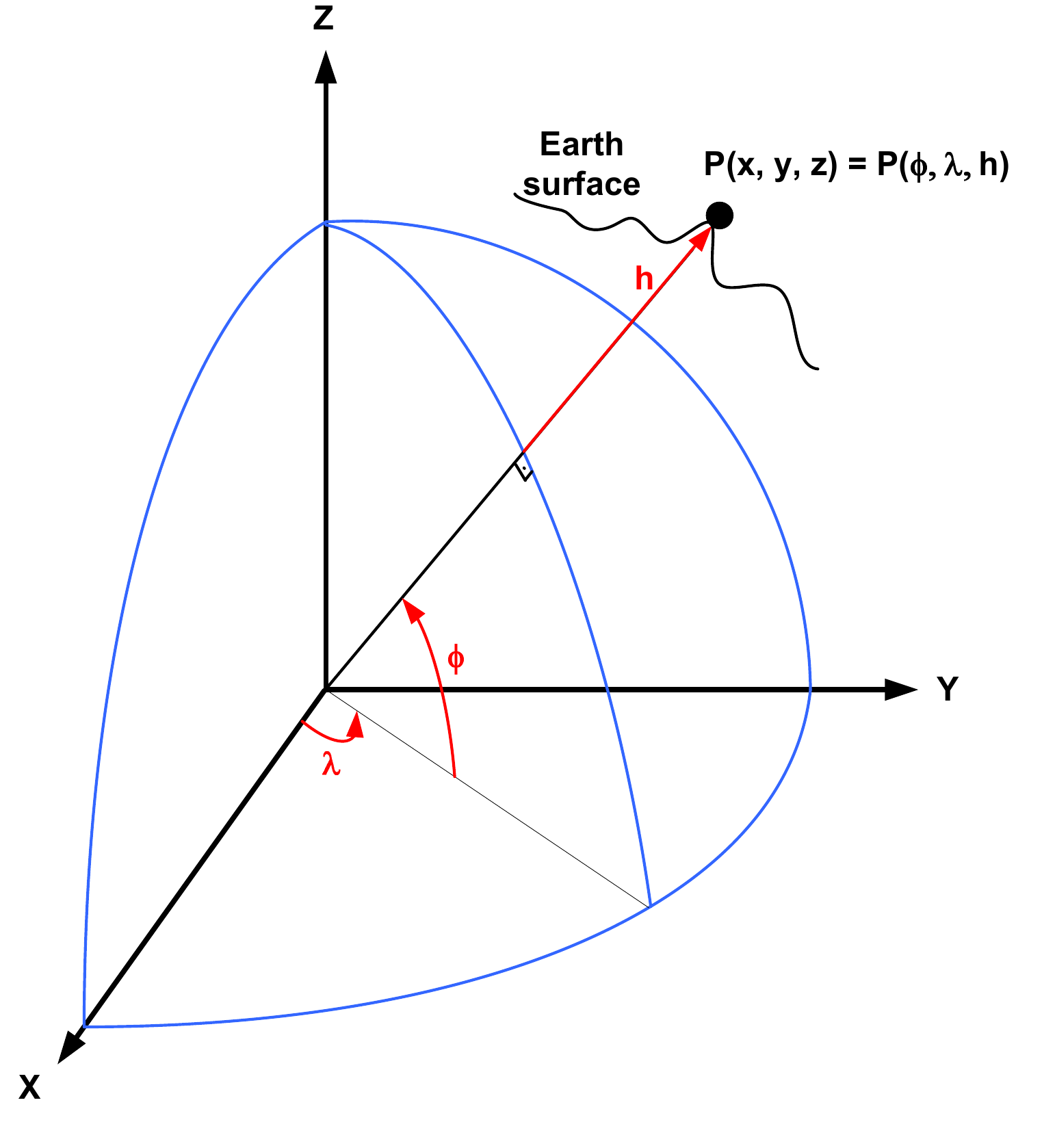}
  \end{center}
  \caption[GPS position parameters]{GPS position parameters in latitude ($\phi$), longitude ($\lambda$) and altitude ($h$) and $x$, $y$ and $z$ in ECEF.  Following~\cite{Stovall1997}.}
  \label{fig:gpsCoordinateSystem}
\end{figure}

The data obtained from the GPS is in well-known NMEA format and includes position, the number of visible satellites and detailed satellite information for a position $P$ on Earth's surface, as shown in Figure~\ref{fig:gpsCoordinateSystem}. 

Using this information, the GPS coordinates can be converted from geodetic latitude ($\phi$), longitude ($\lambda$) and altitude ($h$) notation to ECEF Cartesian coordinates  $\mathbf{x_{gps}}$, $\mathbf{y_{gps}}$ and $\mathbf{z_{gps}}$ as:
\begin{equation}
\begin{array}{l}
\displaystyle x_{gps} = (N+h)\cos(\phi)\cos(\lambda)\\
\displaystyle y_{gps} = (N+h)\cos(\phi)\sin(\lambda)\\
\displaystyle z_{gps} = ((1-e^2)N + h)\sin(\phi)
\end{array} 
\label{eq:ch6latLongAlttoxyz}
\end{equation}
where
\begin{equation}
N = \frac{a}{\sqrt{1.0-e^2\sin(\phi)^2}}
\label{eq:ch6N}
\end{equation}
and $a$ is the WGS84~\cite{Nima2000} ellipsoid constant for equatorial earth radius (6,378,137m), $e^2$ corresponds to the eccentricity of the earth with a value of $6.69437999\times10^{-3}$~\cite{Kaplan2005}. The calculated values form the measurements from the GPS sensor as $m_{gps}=\left( x_{gps},y_{gps},z_{gps}\right)$.

\subsection{IMU position estimate}

Finding the position estimate from the IMU is performed by double-integrating the accelerometer outputs for several samples, the current implementation uses four samples. The first integration, to find the velocity, involves integrating accelerations using $v(t)=v(0)+at$:
\begin{equation}
\begin{array}{l}
\displaystyle v_x=\int_{0}^{T} a_x \mathrm{d}t = v_x(T)-v_x(0)\\
\displaystyle v_y=\int_{0}^{T} a_y \mathrm{d}t = v_y(T)-v_y(0)\\
\displaystyle v_z=\int_{0}^{T} a_z \mathrm{d}t = v_z(T)-v_z(0)
\end{array} 
\label{eq:imuVelocity}
\end{equation}
Since multiple samples are taken, $\mathrm{d}t$ is the time passed for each one of them. The next step is to integrate the velocities from (\ref{eq:imuVelocity}) to find the position using $x(t)=x(0)+vt$ as 
\begin{equation}
\begin{array}{l}
\displaystyle x_{imu}=\int_{0}^{T} v_x \mathrm{d}t = p_x(T)-p_x(0)\\
\displaystyle y_{imu}=\int_{0}^{T} v_y \mathrm{d}t = p_y(T)-p_y(0)\\
\displaystyle z_{imu}=\int_{0}^{T} v_z \mathrm{d}t = p_z(T)-p_z(0)
\end{array} 
\label{eq:imuPosition}
\end{equation}
These calculated positions ($m_{imu}=\left(x_{imu}, y_{imu}, z_{imu}\right)$) are used as the measurements from the IMU.

\section{Fusion Filter}
\label{sec:filter}

The filter designed for integration of the two sensors consists of a state $x$ which includes positional data ($P=(P_x,P_y,P_z)^T$), linear velocities ($V=(V_x,V_y,V_z)^T$):
\begin{equation}
 x= {\left(  P,  V \right)}^{T}
 \label{eq:x}
\end{equation}

A simple state consisting of 6 elements will facilitate obtaining a better performance in speed than one with a larger state. At each iteration, the predict--measure--update cycle of the KF is executed in order to produce a single output from several sensors as the filter output.

In the first stage, \emph{i.e.} prediction, a transition matrix ($F$ of (\ref{eq:F})) is applied to the state $x$ in order to obtain the predicted position:

\begin{equation}
F=\left[
\begin{array}{cccccccccccc}
1 & 0 & 0 & \Delta t & 0 & 0 \\ 
0 & 1 & 0 & 0 & \Delta t & 0 \\ 
0 & 0 & 1 & 0 & 0 & \Delta t \\ 
0 & 0 & 0 & 1 & 0 & 0 \\ 
0 & 0 & 0 & 0 & 1 & 0 \\ 
0 & 0 & 0 & 0 & 0 & 1 
\end{array} 
\right]
\label{eq:F}
\end{equation}
where $\Delta t$ is the time between two prediction stages.

Measurements are obtained from the GPS and the IMU using the values obtained as described in Section~\ref{sec:estimates} and are combined to create a measurement vector:

\begin{equation}
  z = {\left( x_{gps} + x_{imu},  y_{gps} + y_{imu}, z_{gps} + z_{imu} \right)}^{T}
  \label{eq:measurement}
\end{equation}

Here, the IMU measurements for position are used as offsets to the position obtained
from the most recent GPS fix.

\section{DFA based Model Transitions}
\label{sec:transitions}

The difference between the measurements ($z$) and the prediction ($h\hat{x}$), omitting the subscripts indicating time, is defined as the innovation ($y$):
\begin{equation}
  y = z - h\hat{x}
  \label{eq:innovation}
\end{equation}

The innovation vector has 3 components for position elements as $y_x$, $y_y$ and $y_z$. The DFA model presented here uses the magnitude of these to define the filter divergence as 
\begin{equation}
  I = \sqrt{y_x^2 + y_y^2 + y_z^2}
  \label{eq:I}
\end{equation}
and uses the following rules to assign the values of $I$ into different classes named $I_0$, $I_1$ and $I_2$ which are defined as

\begin{equation}
I=\left\{
\begin{array}{cc}
I_0: & I < 3.0 \\ 
I_1: & 3.0 \le I < 7.5 \\ 
I_2: & 7.5 \le I \\ 
\end{array} 
\right.
\end{equation}

\begin{figure}[h!t]
  \begin{center}
    \includegraphics[width=\1]{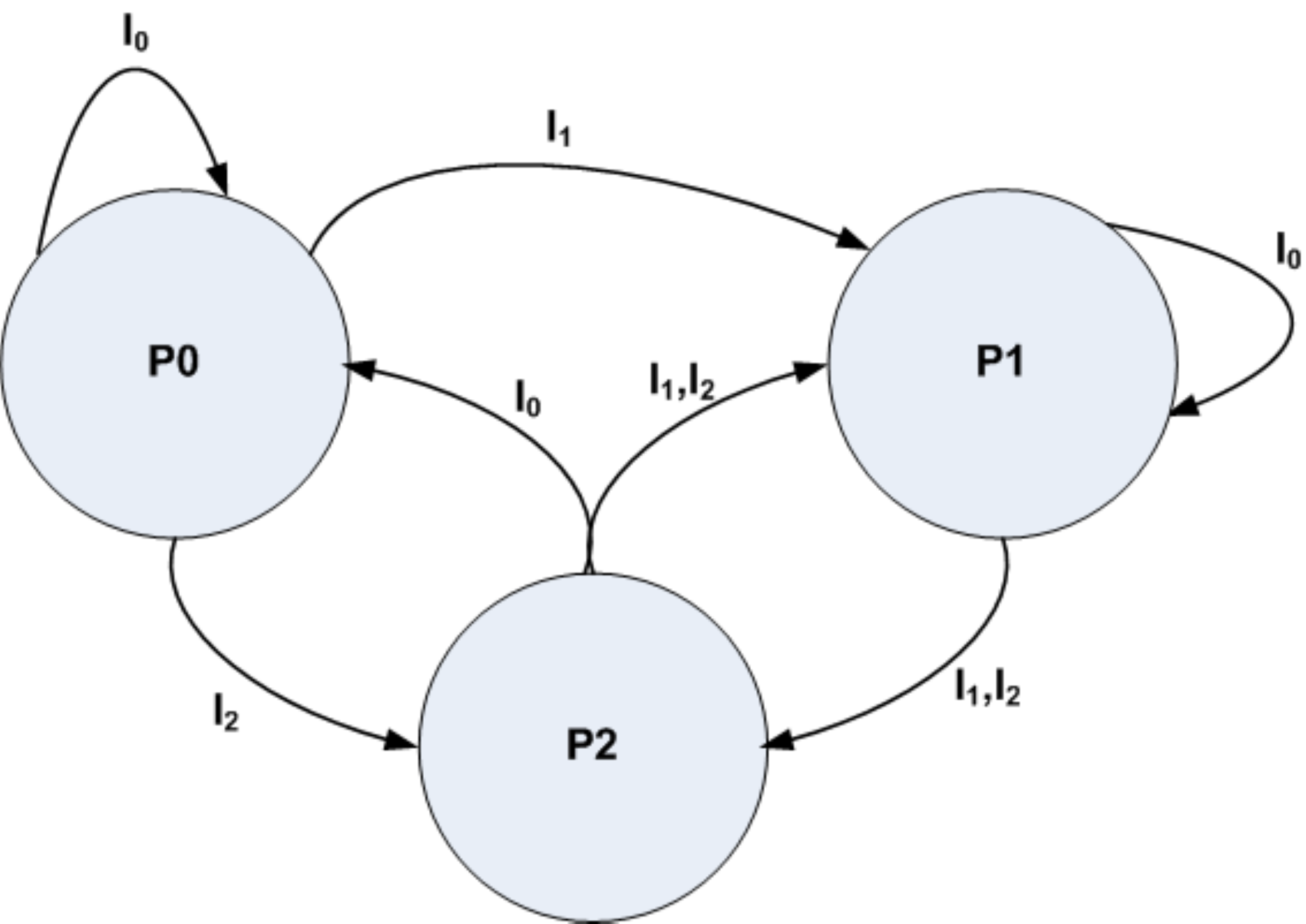}
  \end{center}
  \caption[]{DFA model for the model transitions}
  \label{fig:models}
\end{figure}

A DFA consists of several elements which can be listed as states, input symbols and transition rules~\cite{Hopcroft2007, Uchida2011}. The states of the DFA defined in Figure~\ref{fig:models} correspond to different motion models.

The classes ($I_0$, $I_1$ and $I_2$) are considered as the input symbols used for the DFA model. Finally, the transitions between states model the selection mechanism presented in this paper. 

When a model ($P_i$) is selected, the value for $i$ is used as a velocity coefficient ($c_i$) in the transition function ($F$) for position ($\hat{x}_{P}=x_P+c_{i}V\Delta t$). For instance,
\texttt{P0} indicates a stationary transition model where the current values
for position (\texttt{P}) will be unchanged in the
predicted state, whereas \texttt{P2} indicates a motion model where position
is predicted with twice the current positional velocities ($\hat{x}_{P}=x_P+(2\times V)\Delta t$) in order to adapt any sudden changes in the estimated position.

\begin{figure}[h!t!p]
  \begin{center}
    \includegraphics[width=\1]{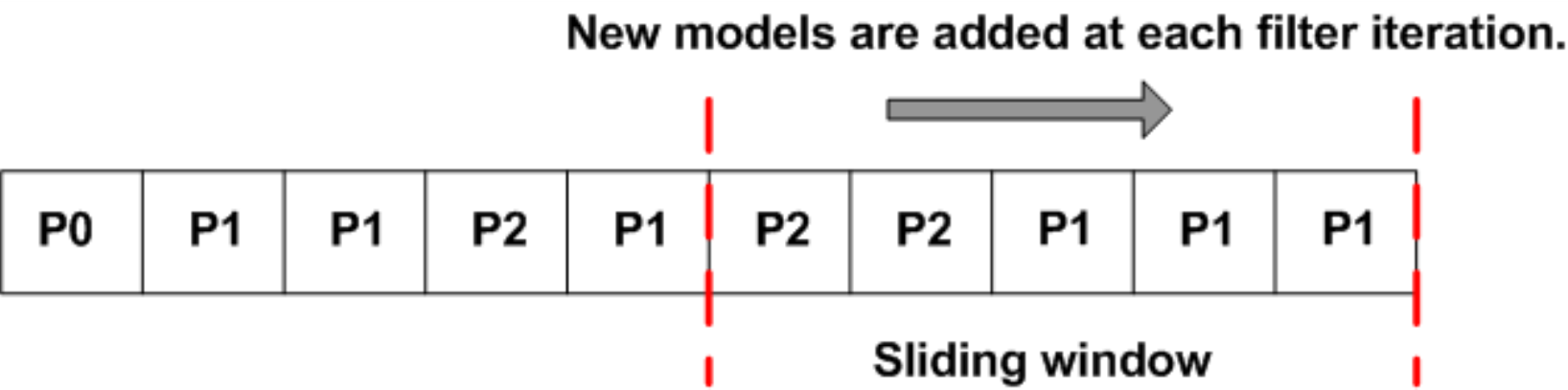}
  \end{center}
  \caption[]{Sliding window for preventing frequent model transitions}
  \label{fig:slidingWindow}
\end{figure}

During experiments it was observed that in some cases selected models could be changing very often. A sliding window filter was applied to the results of the model selection logic in order to prevent frequent transitions between different motion models. In the implementation, the most recent five models were averaged to obtain the final motion model as illustrated in Figure~\ref{fig:slidingWindow}.

\section{Results}
\label{sec:results}
Experiments were conducted using low-cost GPS and IMU sensors mounted on a
cycle helmet for a user walking with varying speed. Sampling rate for the IMU was selected as 20 milliseconds and a GPS fix was received every second.

\begin{figure}[h!t!p!b]
  \begin{center}
    \subfigure[]{\includegraphics[width=\1]{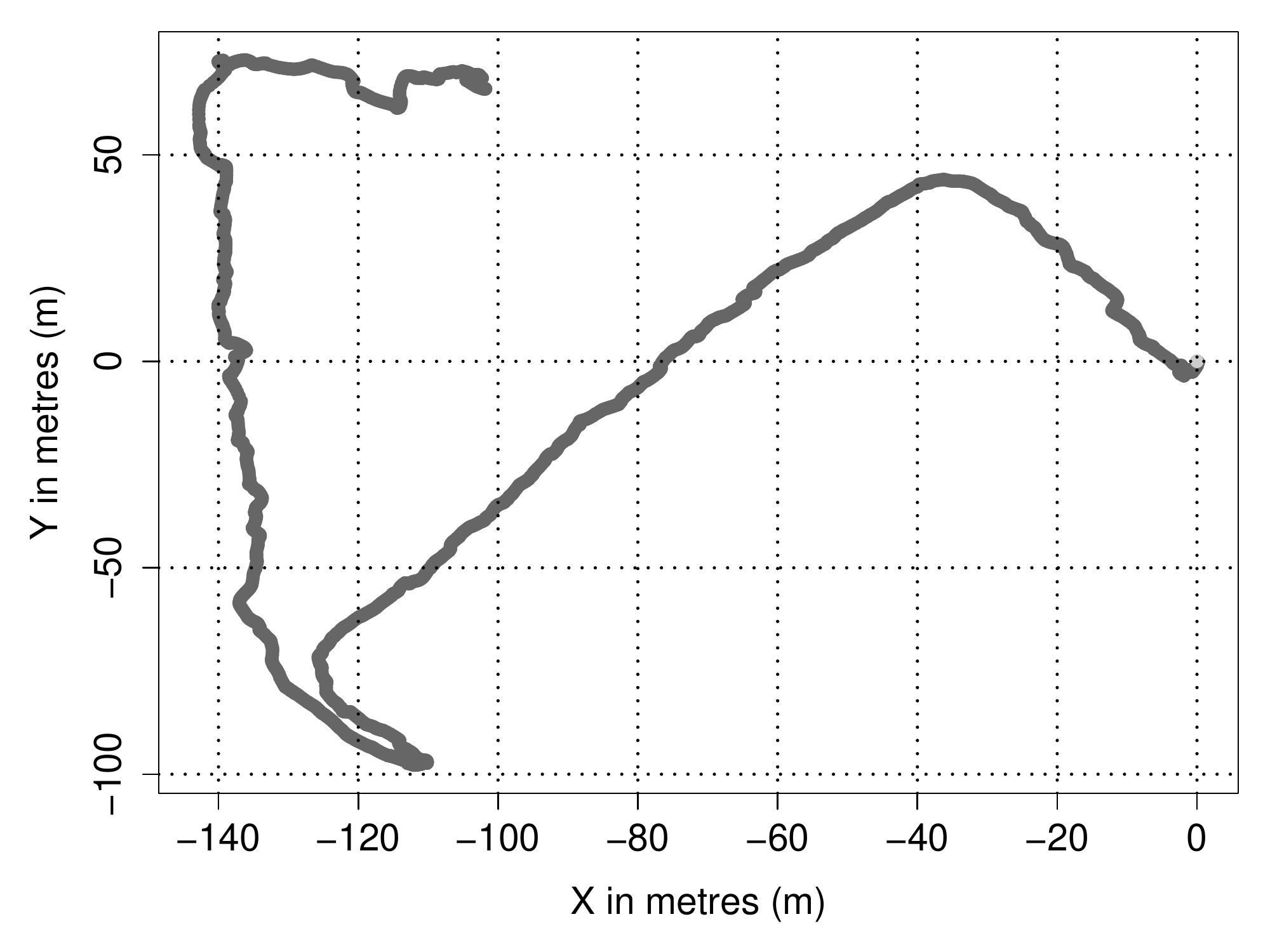}}\\
    \subfigure[]{\includegraphics[width=\1]{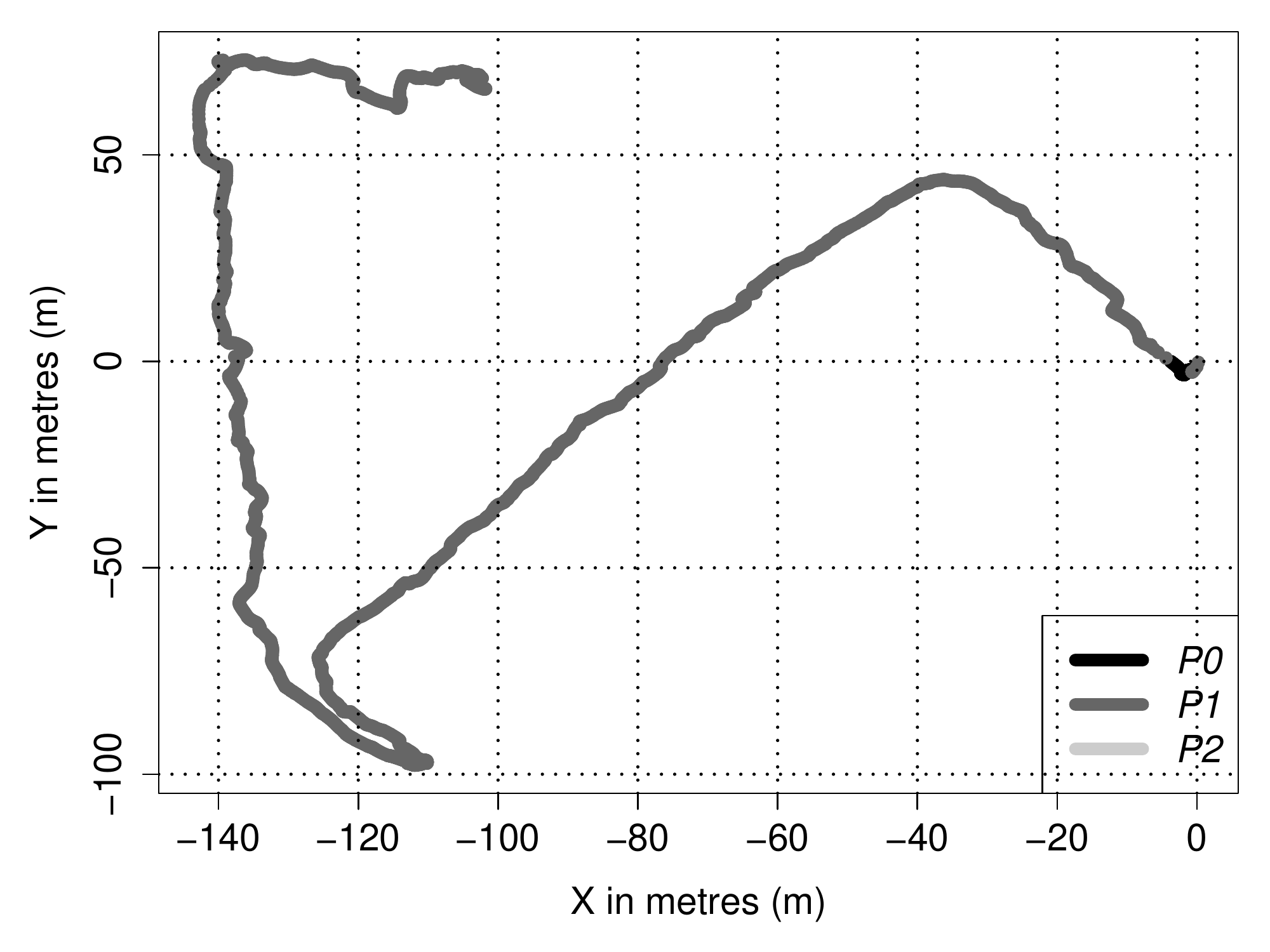}}    
  \end{center}
  \caption[]{Trajectory results for Dataset 1. (a) Static motion model (b) DFA models}
  \label{fig:trajectoryDataset1}
\end{figure}

\begin{figure}[h!t!p!b]
  \begin{center}
    \subfigure[]{\includegraphics[width=\1]{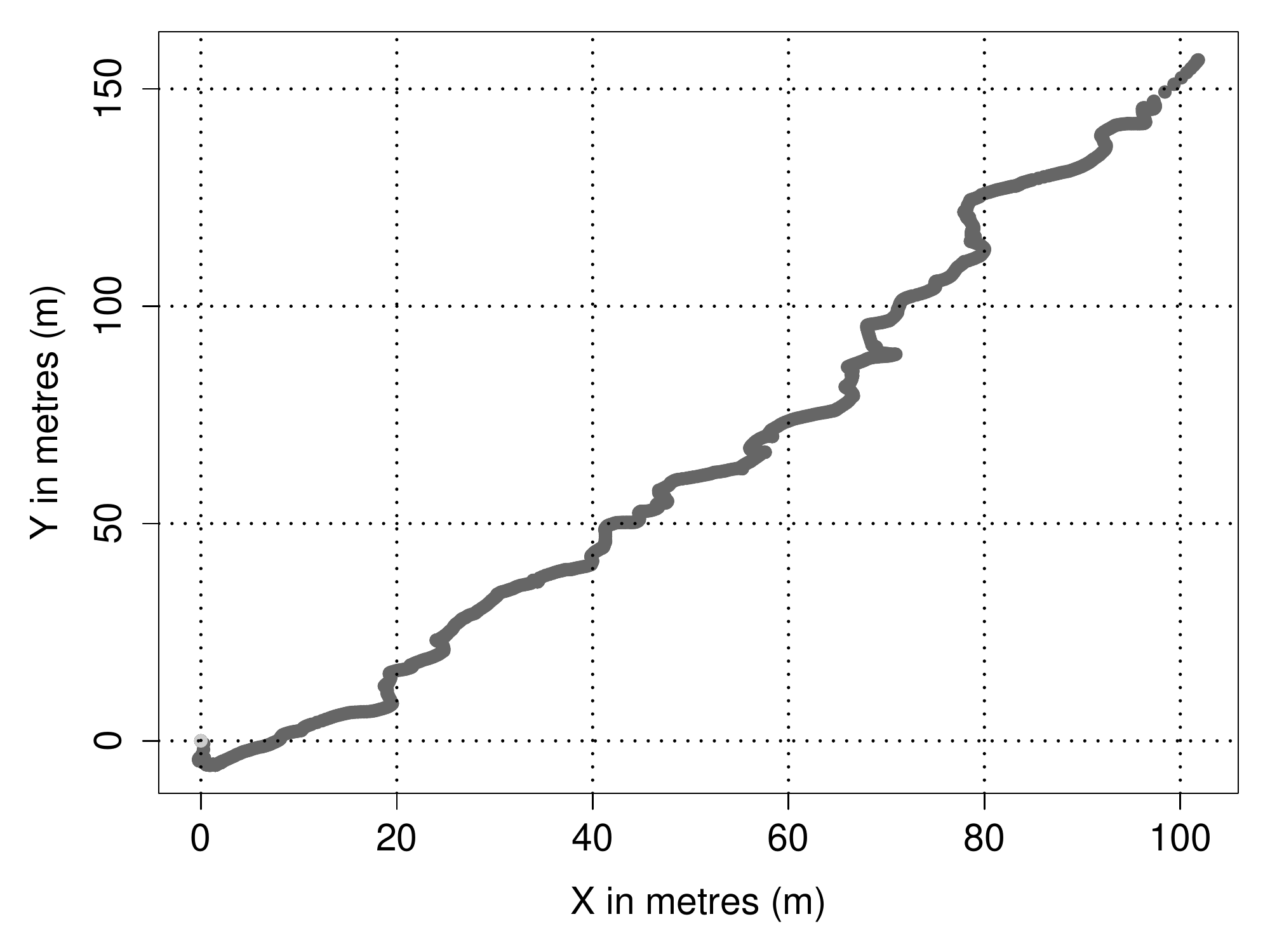}}\\
    \subfigure[]{\includegraphics[width=\1]{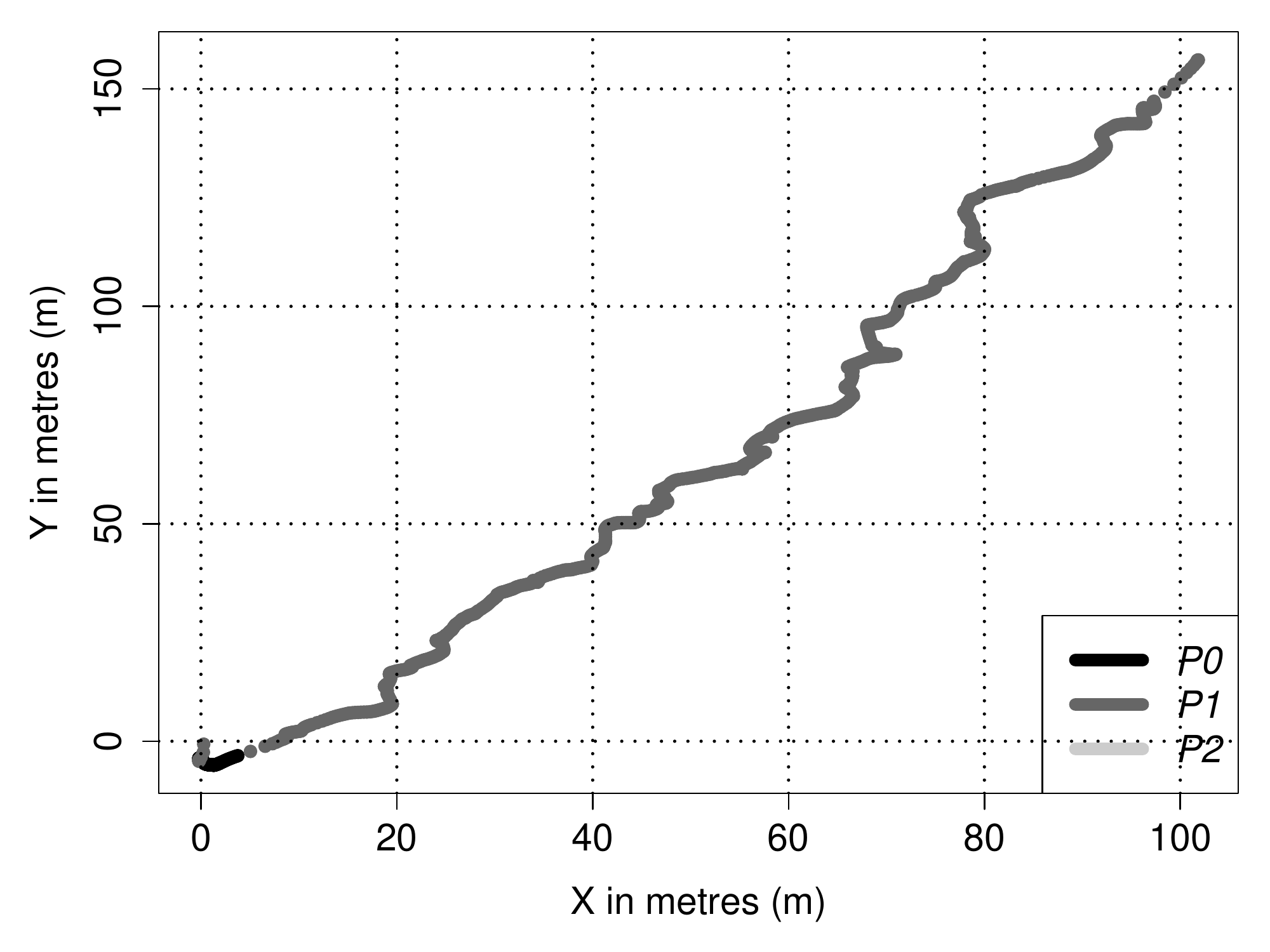}}    
  \end{center}
  \caption[]{Trajectory results for Dataset 2. (a) Static motion model (b) DFA models}
  \label{fig:trajectoryDataset2}
\end{figure}

\begin{figure}[h!t!p!b]
  \begin{center}
    \subfigure[]{\includegraphics[width=\1]{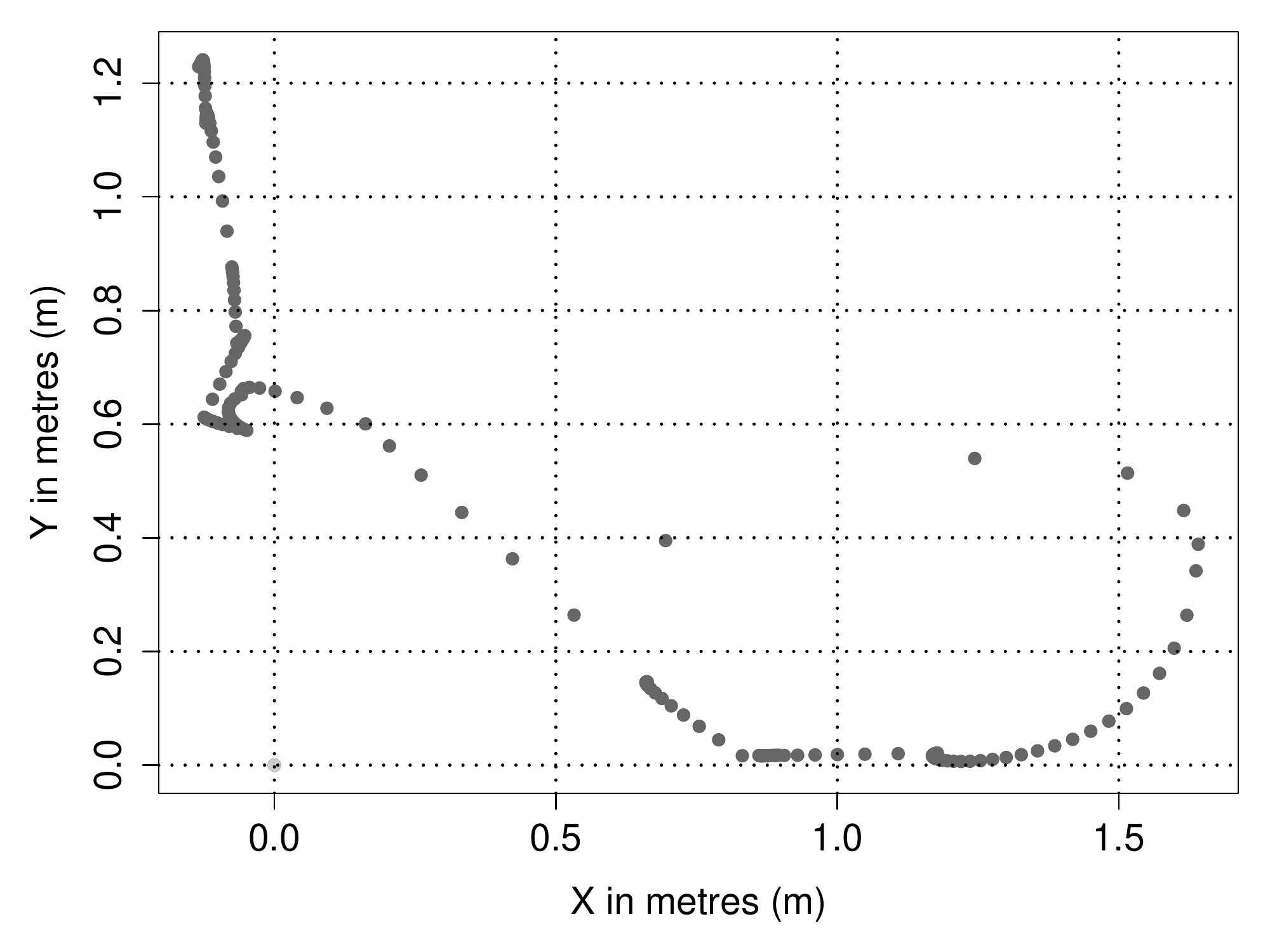}}\\
    \subfigure[]{\includegraphics[width=\1]{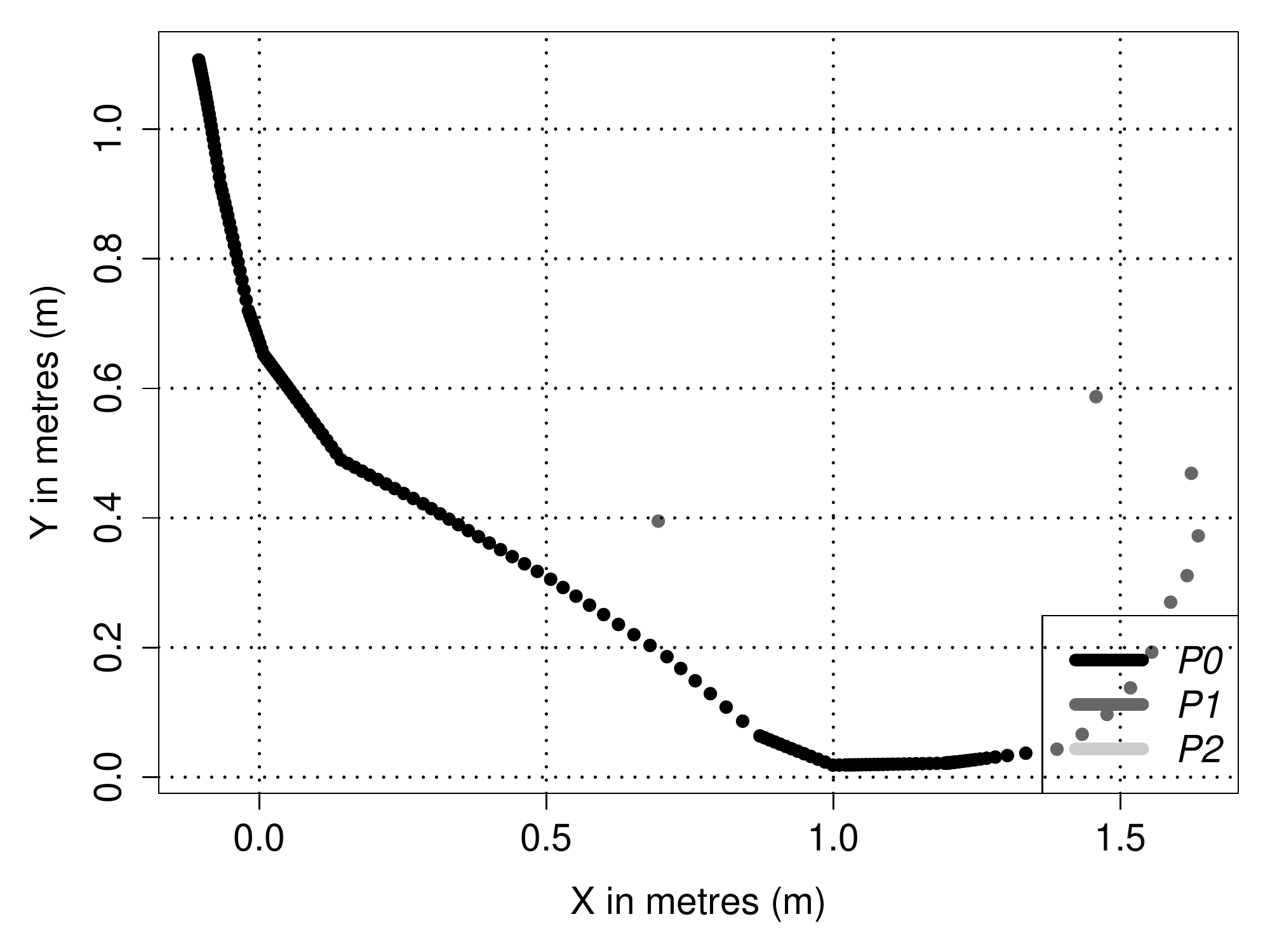}}    
  \end{center}
  \caption[]{Trajectory results for Dataset 3. (a) Static motion model (b) DFA models}
  \label{fig:trajectoryDataset3}
\end{figure}

\begin{figure}[h!t!p!b]
  \begin{center}
    \subfigure[]{\includegraphics[width=\1]{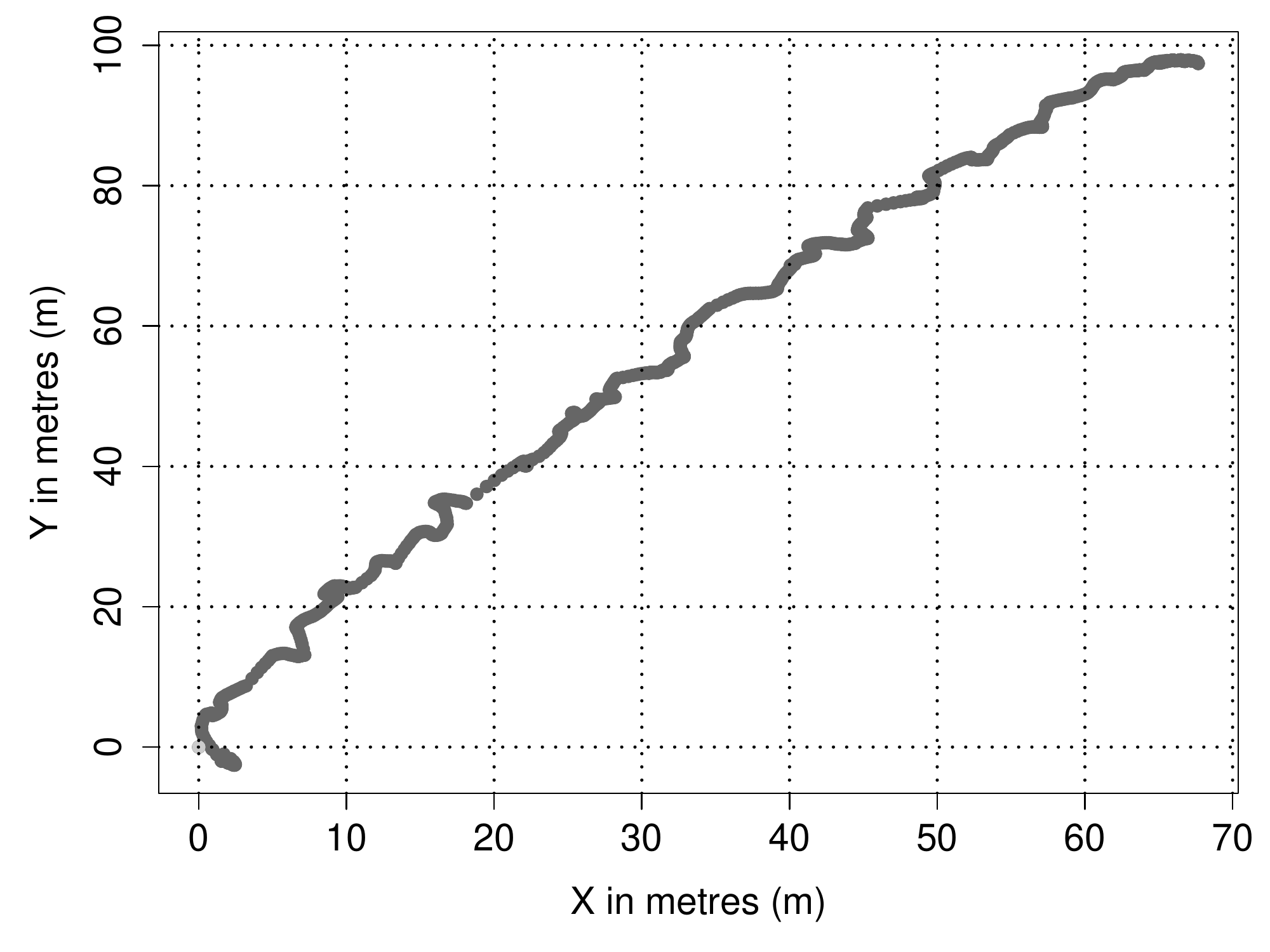}}\\
    \subfigure[]{\includegraphics[width=\1]{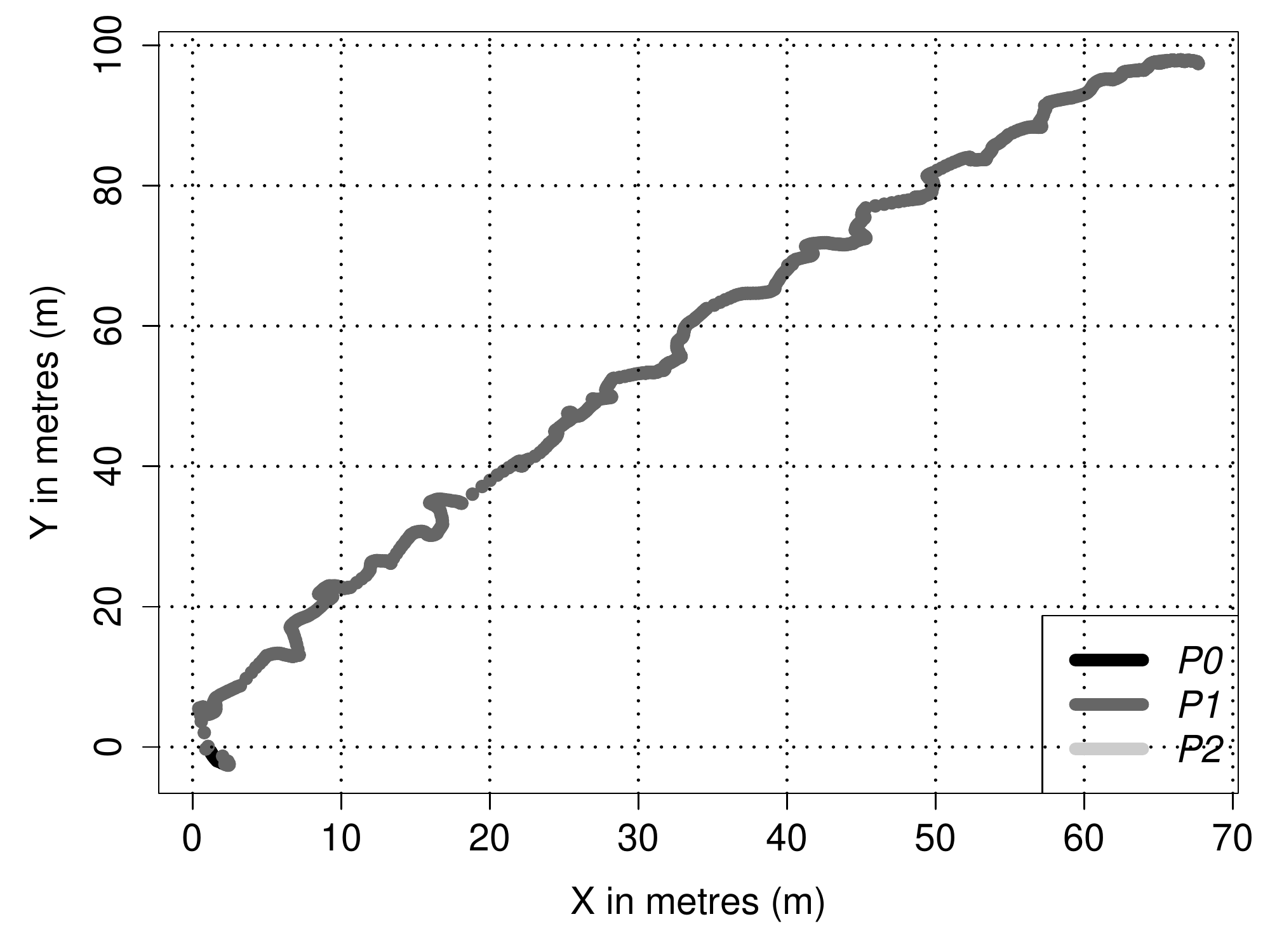}}    
  \end{center}
  \caption[]{Trajectory results for Dataset 4. (a) Static motion model (b) DFA models}
  \label{fig:trajectoryDataset4}
\end{figure}

\begin{figure}[h!t!p!b]
  \begin{center}
    \subfigure[]{\includegraphics[width=\1]{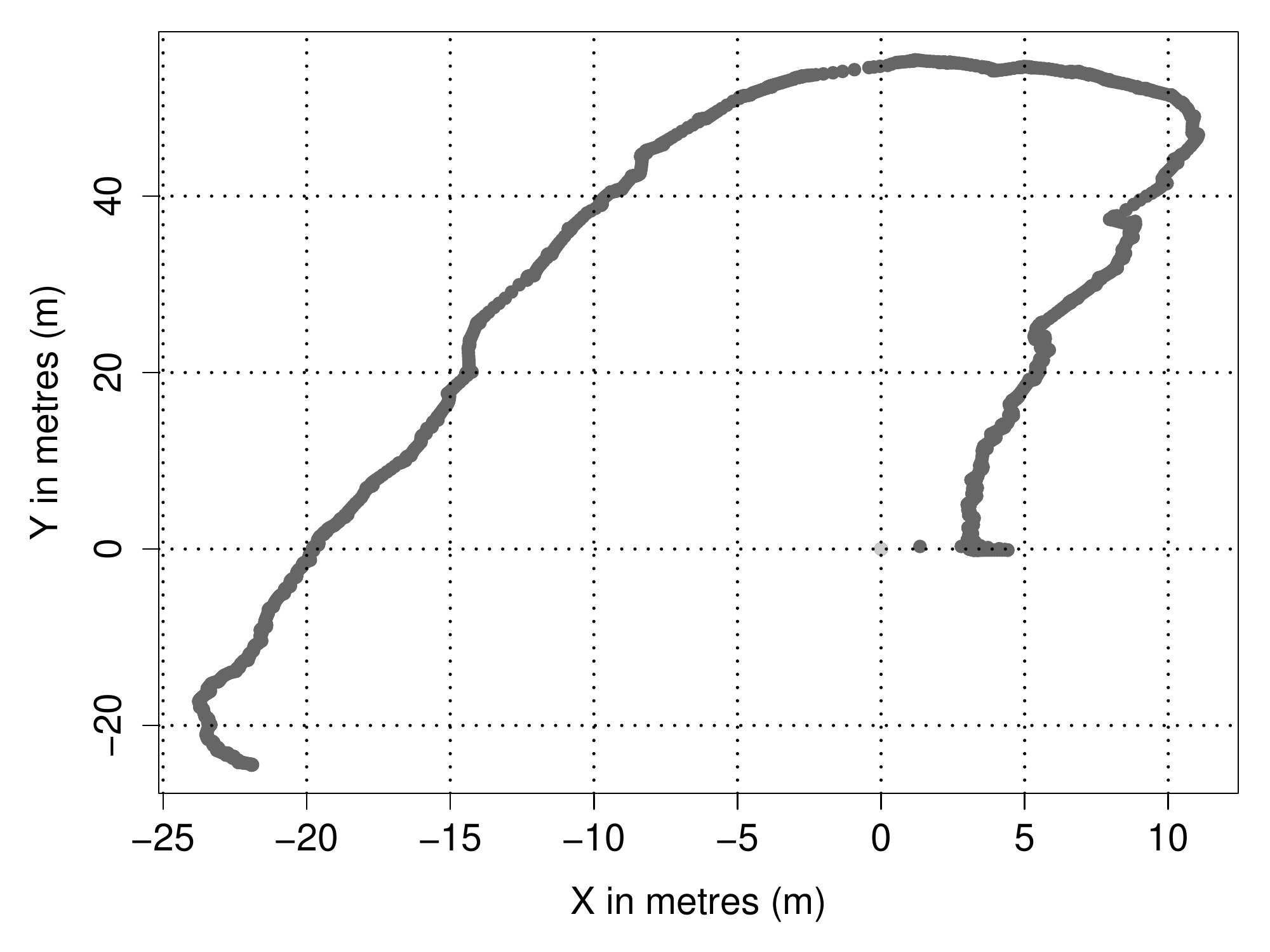}}\\
    \subfigure[]{\includegraphics[width=\1]{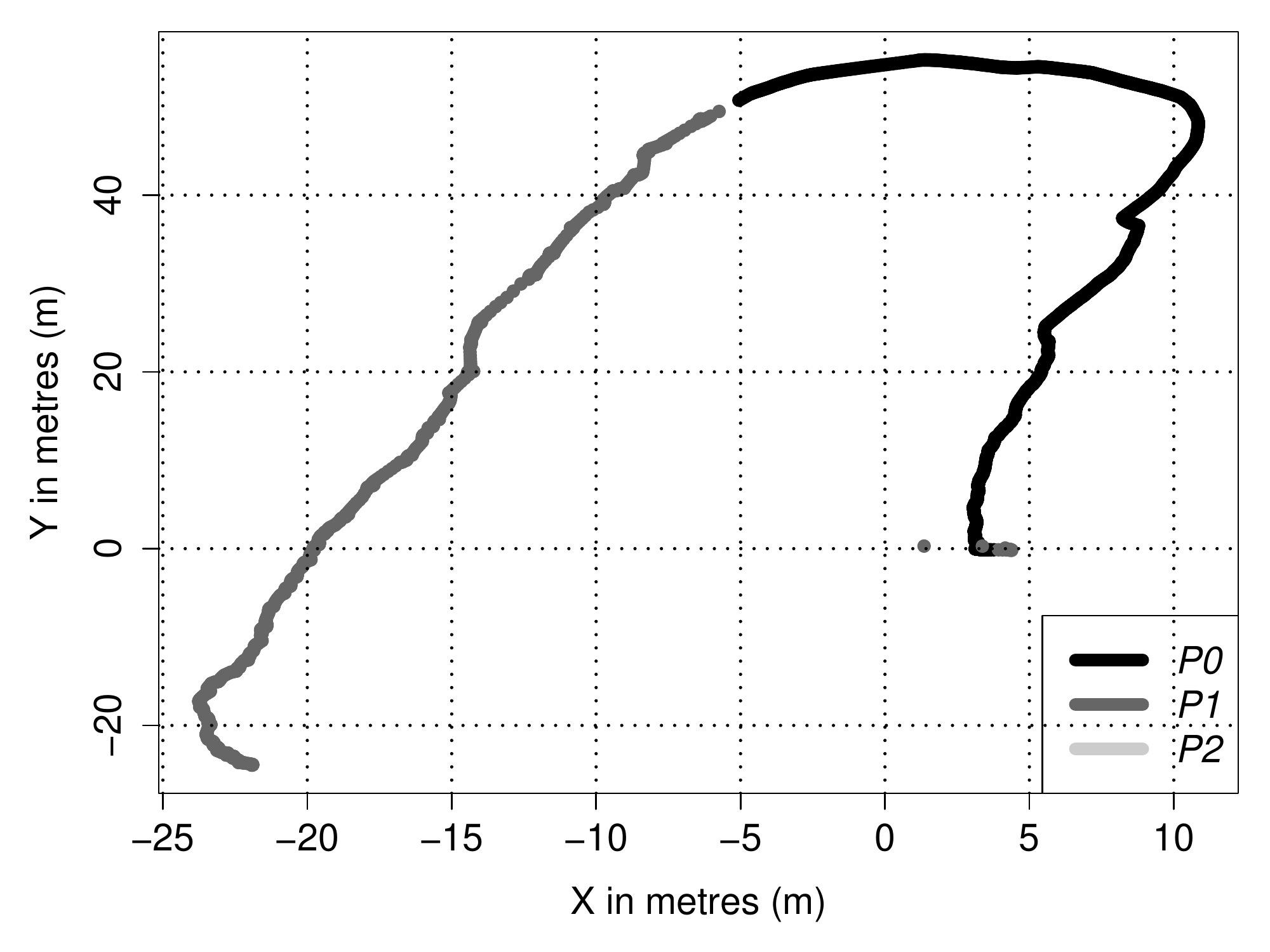}}    
  \end{center}
  \caption[]{Trajectory results for Dataset 5. (a) Static motion model (b) DFA models}
  \label{fig:trajectoryDataset5}
\end{figure}

Figures~\ref{fig:trajectoryDataset1} to~\ref{fig:trajectoryDataset5} show the estimated paths using integration of the two sensors and employing different motion models. Portions of the estimated paths are coloured differently in order to indicate the type of the motion model used for estimation. It is important to note that the static model used in the results correspond to \texttt{P1} and hence is drawn in the same colour.

Dataset 3  was acquired while the sensors were completely stationary, the accuracy of the sensor fusion is found in this case as 1.5m which is, indeed, less than the accuracy of the GPS used in the experiments (given as 2.5m in the product specification) --- a benefit of sensor fusion. The DFA model selection logic reduced this error even further since the motion model is correctly recognized as \texttt{P0} (see Figure~\ref{fig:trajectoryDataset3}).

\begin{figure}[h!t]
  \begin{center}
    \includegraphics[width=\1]{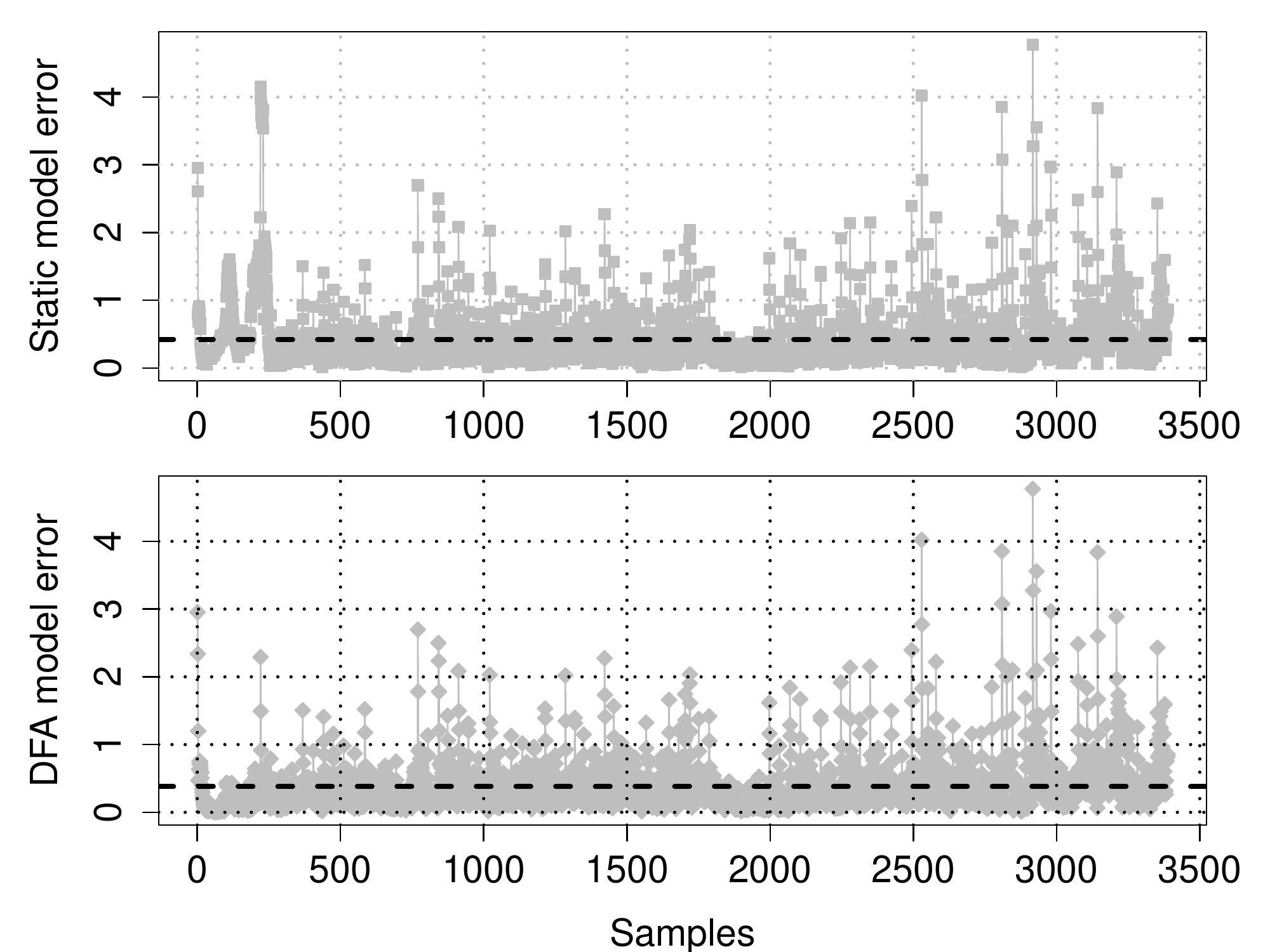}
  \end{center}
  \caption[]{Filter errors for Dataset 1}
  \label{fig:errorsDataset1}
\end{figure}

\begin{figure}[h!t!p!b]
  \begin{center}
    \includegraphics[width=\1]{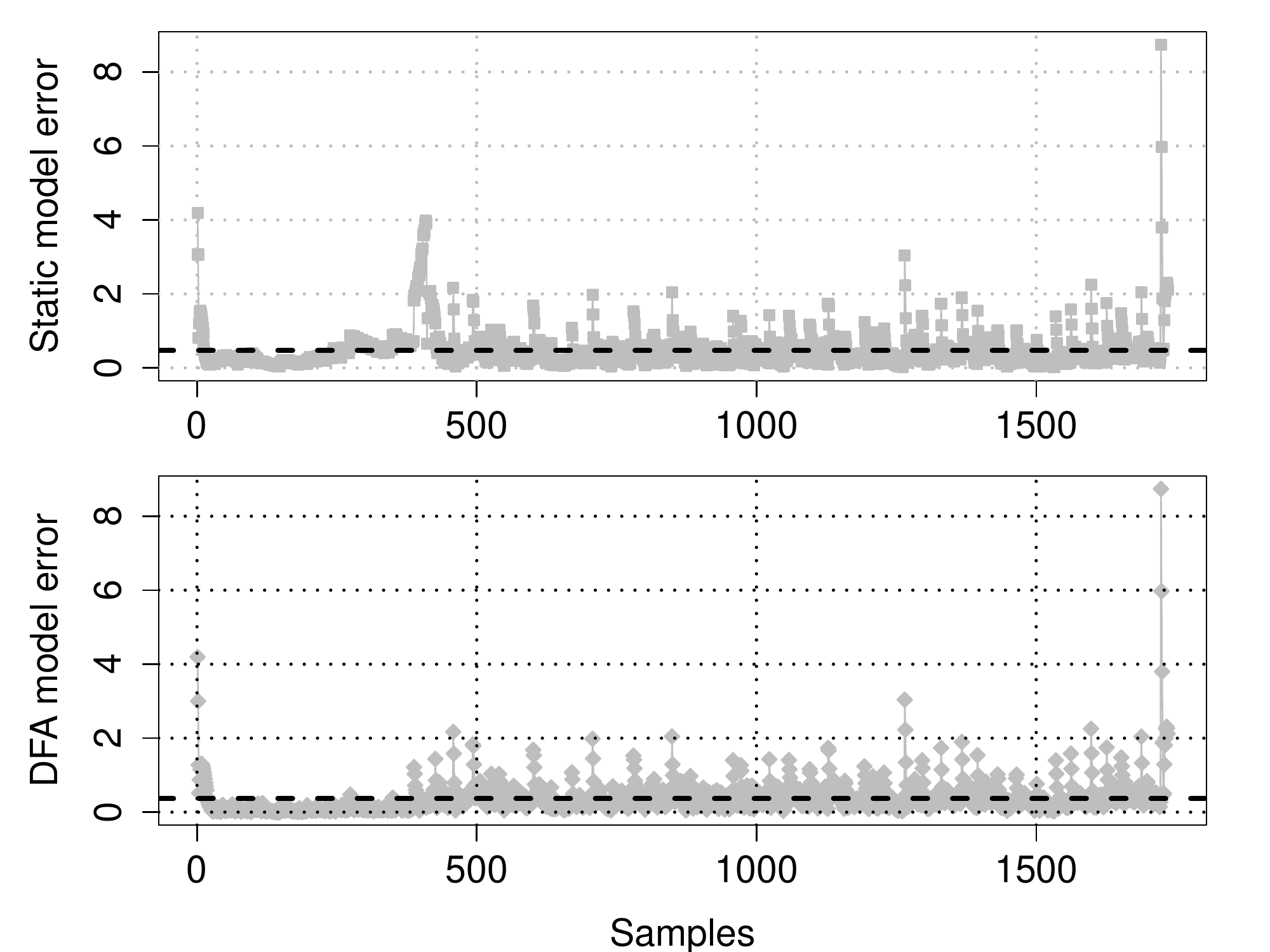}
  \end{center}
  \caption[]{Filter errors for Dataset 2}
  \label{fig:errorsDataset2}
\end{figure}

\begin{figure}[h!t!p!b]
  \begin{center}
    \includegraphics[width=\1]{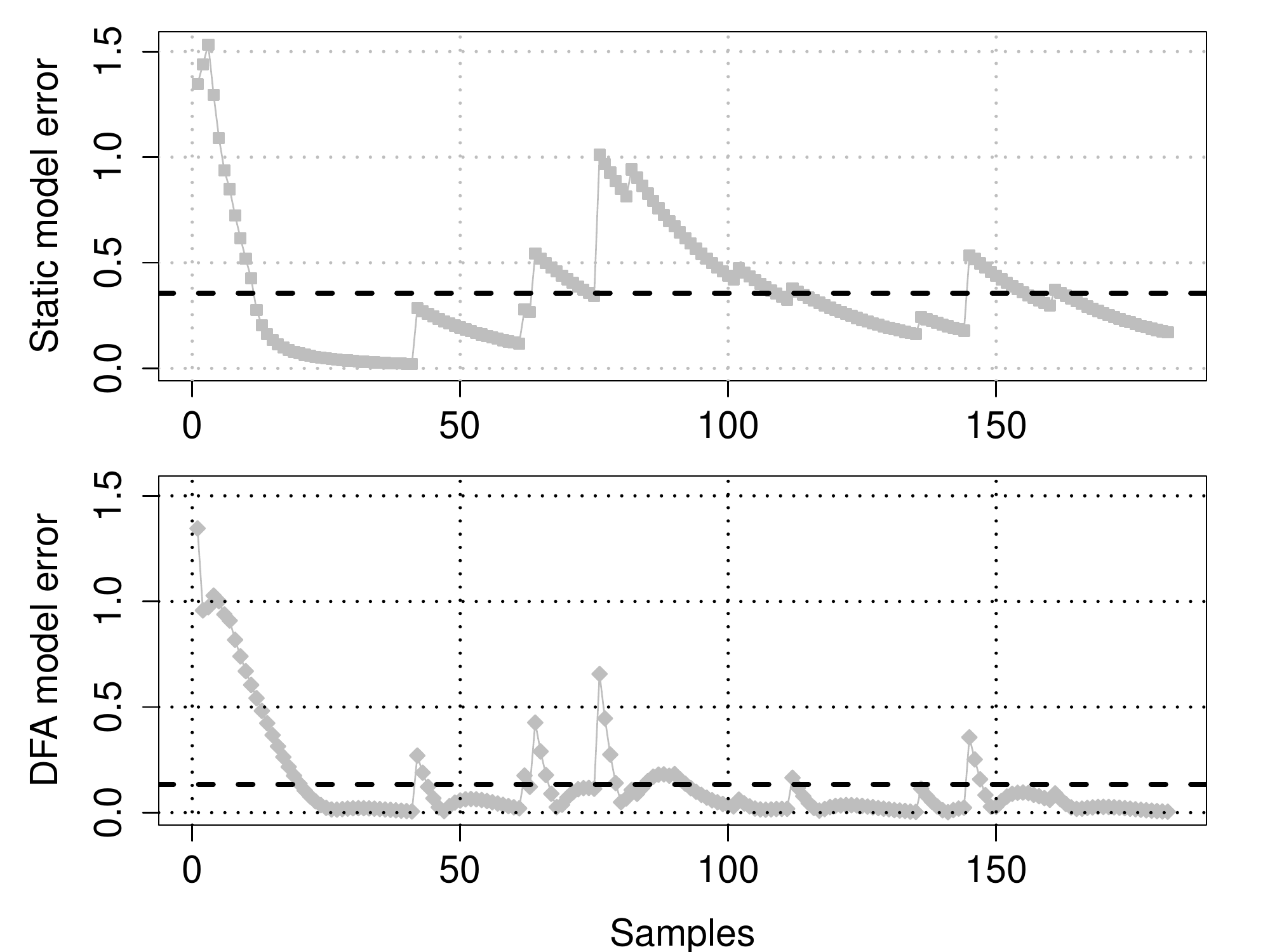}
  \end{center}
  \caption[]{Filter errors for Dataset 3}
  \label{fig:errorsDataset3}
\end{figure}

\begin{figure}[h!t!p!b]
  \begin{center}
    \includegraphics[width=\1]{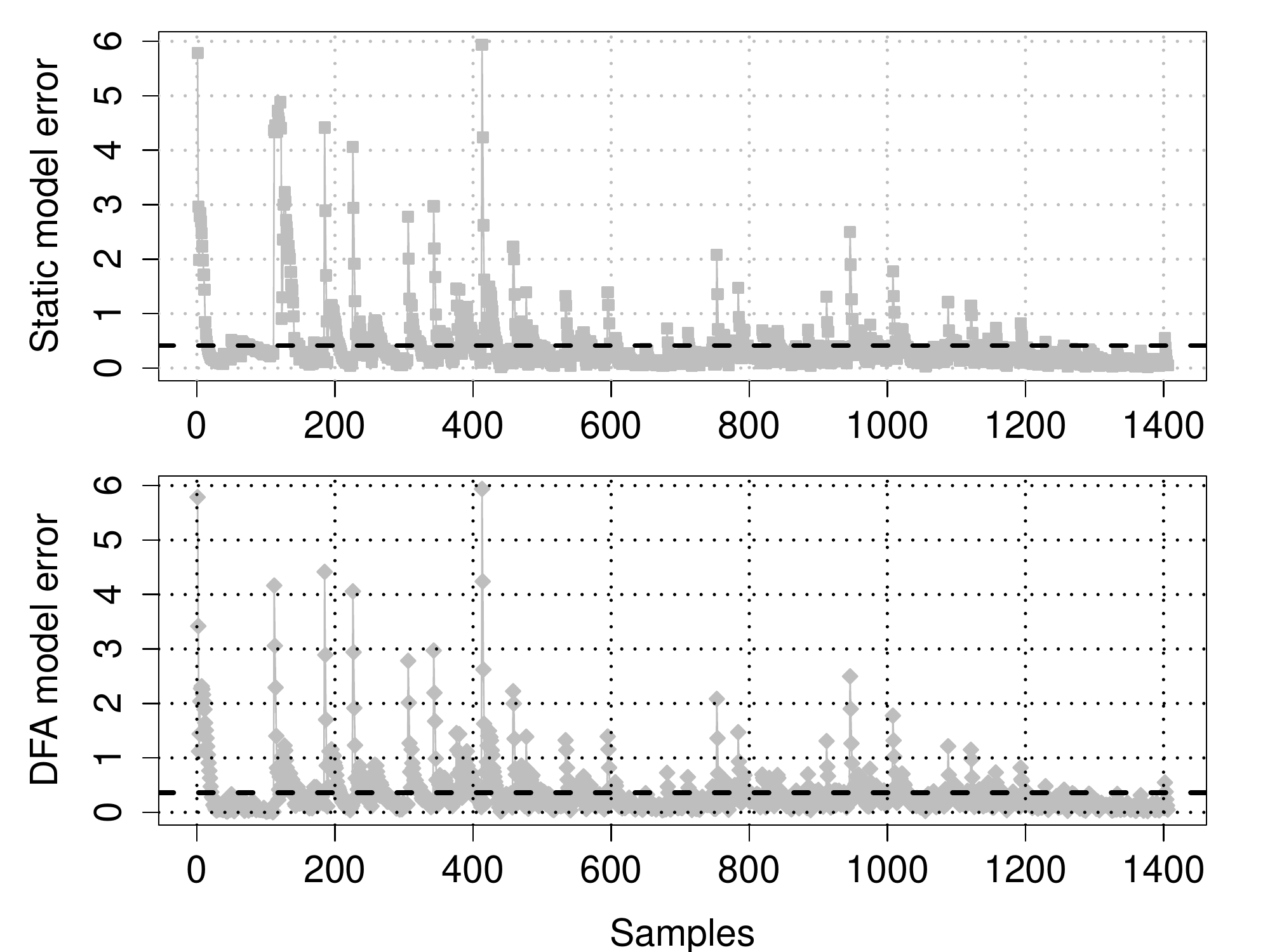}
  \end{center}
  \caption[]{Filter errors for Dataset 4}
  \label{fig:errorsDataset4}
\end{figure}

\begin{figure}[h!t!p!b]
  \begin{center}
    \includegraphics[width=\1]{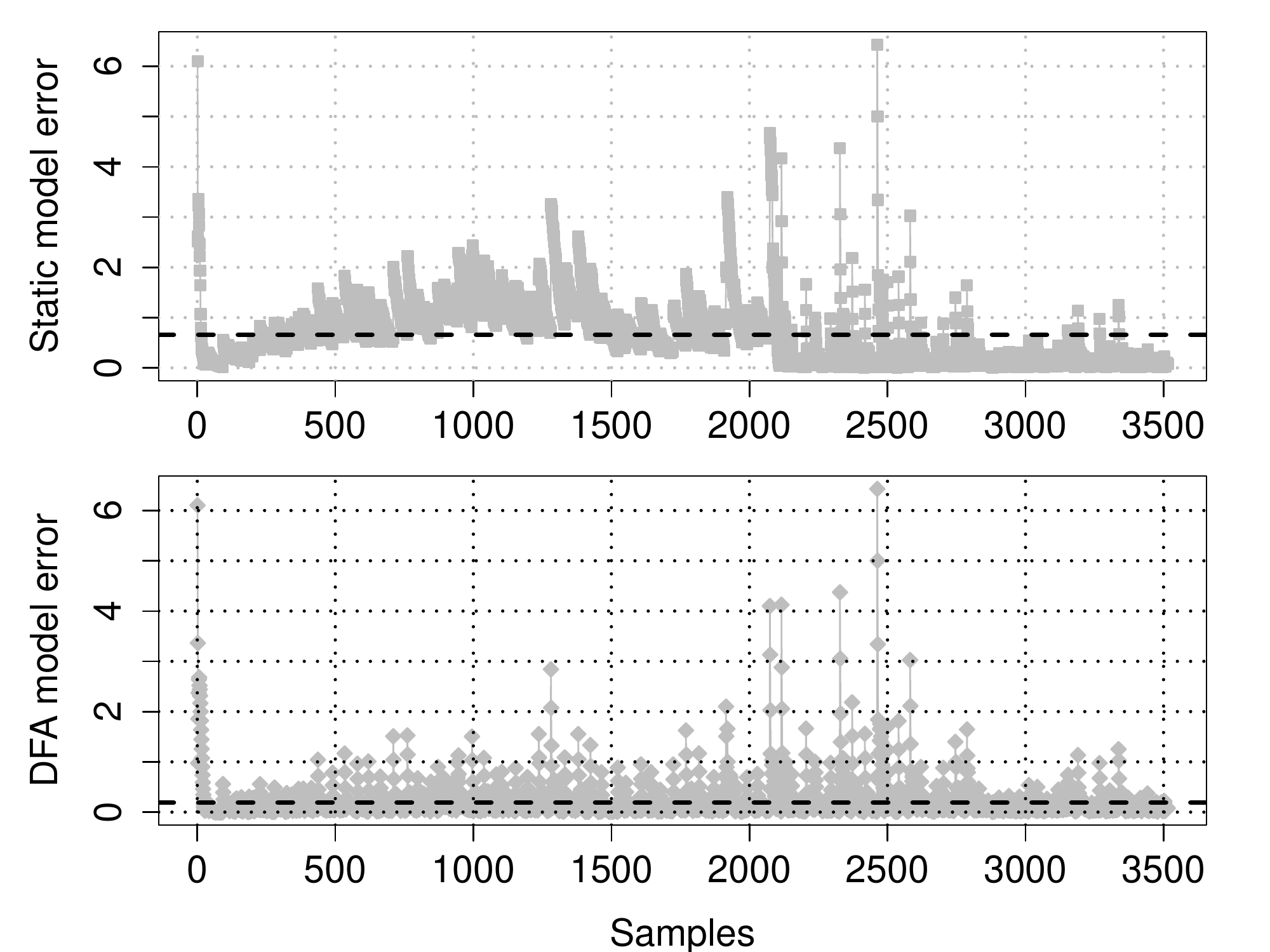}
  \end{center}
  \caption[]{Filter errors for Dataset 5}
  \label{fig:errorsDataset5}
\end{figure}

Filter errors are presented in Figures~\ref{fig:errorsDataset1} to~\ref{fig:errorsDataset5}. Note that these errors are an indicator of the difference between the filter predictions and the actual values of the measurements. It can be seen that the filter error is reduced when the DFA models are employed.

\clearpage

\clearpage

\section{An AR Game -- Treasure Hunt}
\label{sec:arGame}

This section presents an AR game, which works using the DFA based sensor fusion algorithm described earlier in the paper. The aim of the game is to collect items and direct the user to test the accuracy of the approach, albeit unconsciously. 

\begin{figure}[h!t!p]
  \begin{center}
    \subfigure[Chest]%
    {\includegraphics[width=\2]{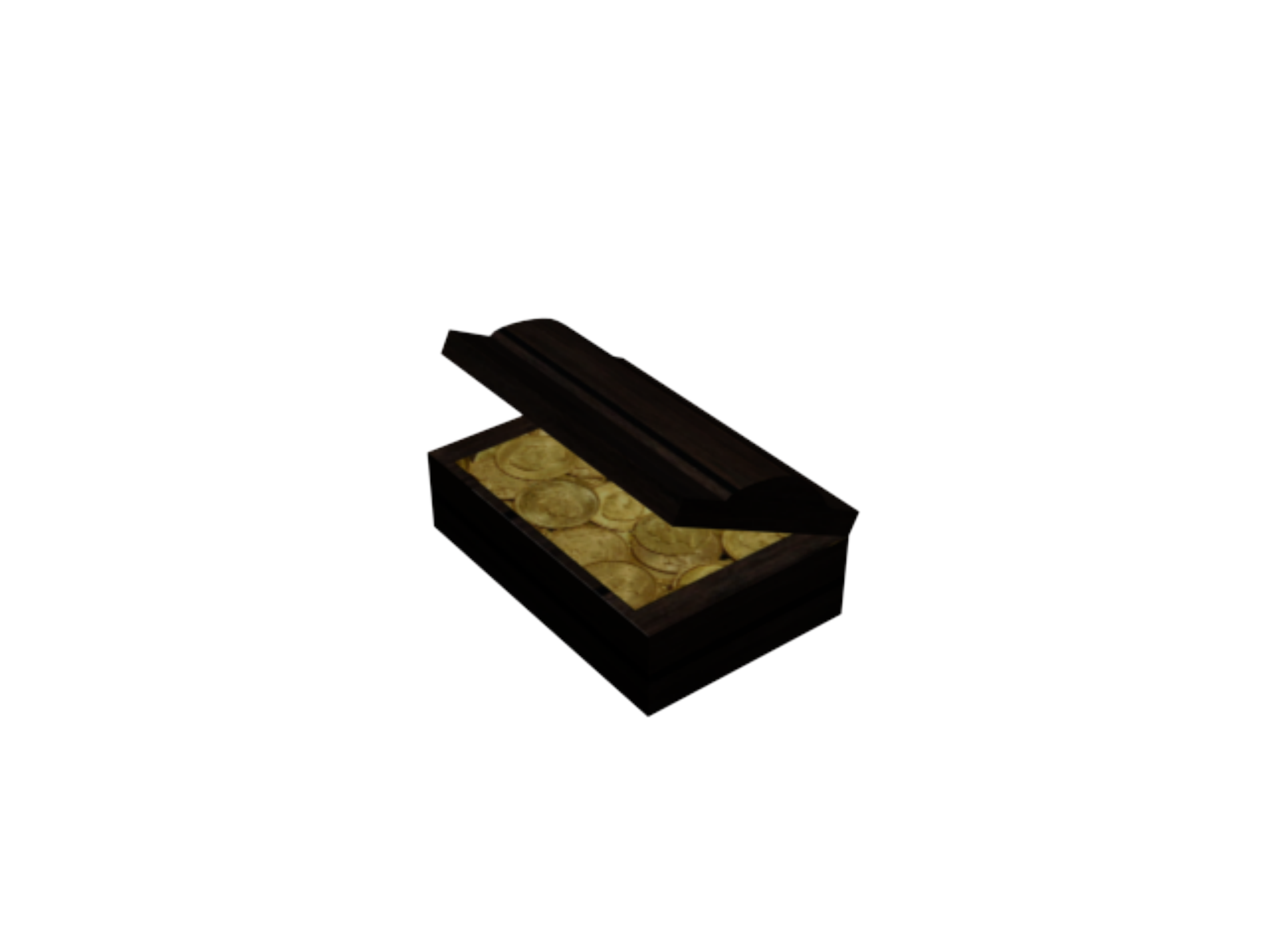}}
    \subfigure[Coin]%
    {\includegraphics[width=\2]{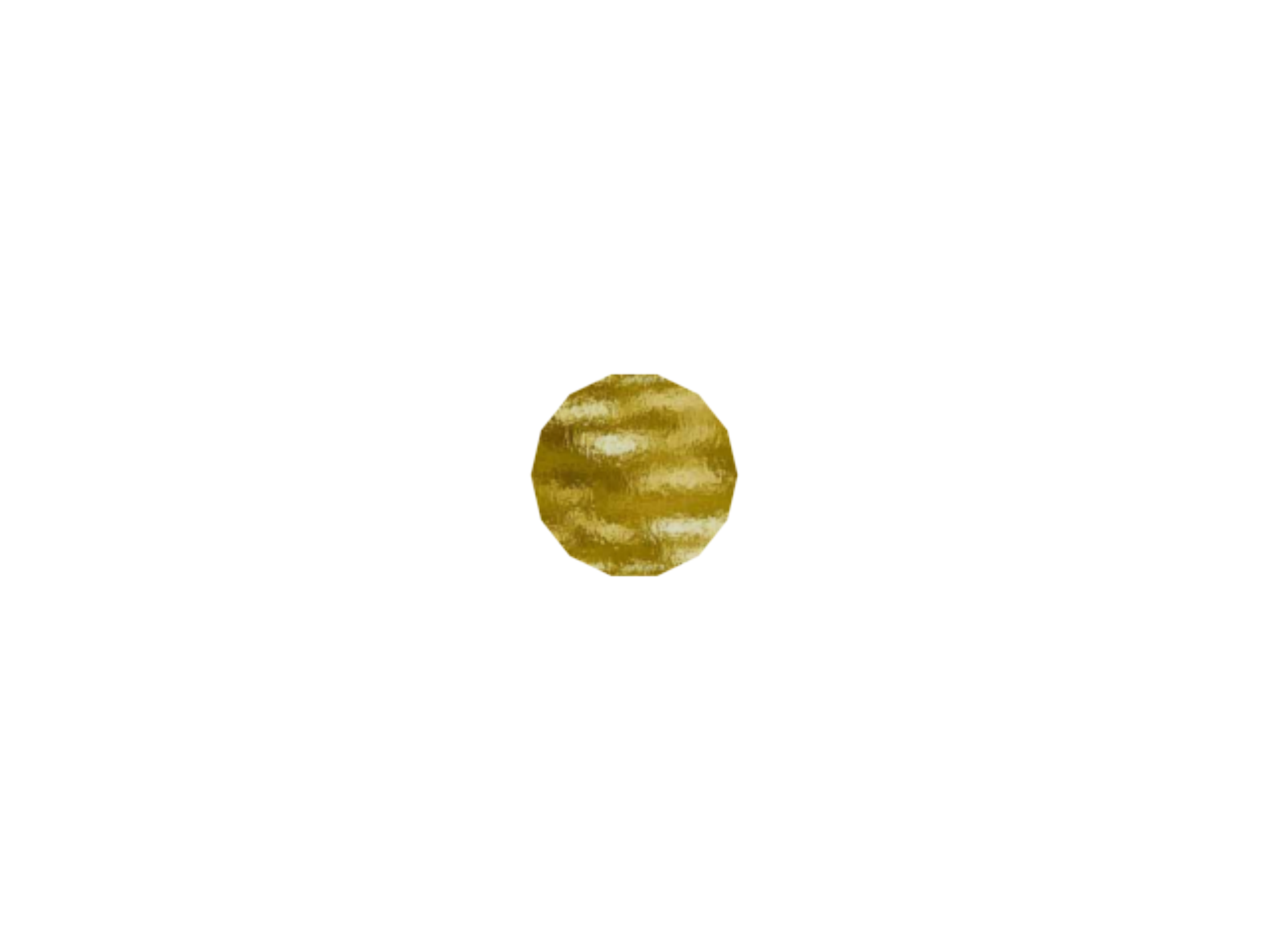}}   
  \end{center}
  \caption[Models used in the AR game]{Models used in the AR game}
  \label{fig:ch7gameModels}
\end{figure}

The game presents an egocentric view of the environment, as in First Person Shooter (FPS) games. The rules of the game are quite simple: the user needs to reach and collect all the reward items available as quickly as possible. When he or she reaches an item, the score is incremented by an amount that depends on the type of item encountered. The game provides three types of items: small coins, large coins and a chest (Figure~\ref{fig:ch7gameModels}), with rewards of 10, 30 and 50 points respectively.

After the game is initialized with the positions of all items set, the game loop starts. The coin models use the animator and they rotate about their axes while the chest models remain static. 

At each frame, the position of the user is checked against the item positions by calculating the distance between them. If this distance is less than some threshold value (done so that there is some tolerance against positioning inaccuracies), then the score is updated, the item is set as `hit' and a sound file is played. The items collected by the user simply disappear. A timer is used for two purposes. First, it is constantly updated in the display to provide feedback to the user. It is also used to decay the score
\begin{equation}
  score = initialScore \times c/time
  \label{eq:timing}
\end{equation}
where $score$ is the final score to be added and $initialScore$ correspond to the rewards mentioned above. The constant $c$ is selected as $5.0$ arbitrarily. This forces the user to collect the game tokens quickly.

\begin{figure}[h!t!p]
  \begin{center}
    \includegraphics[width=\1]{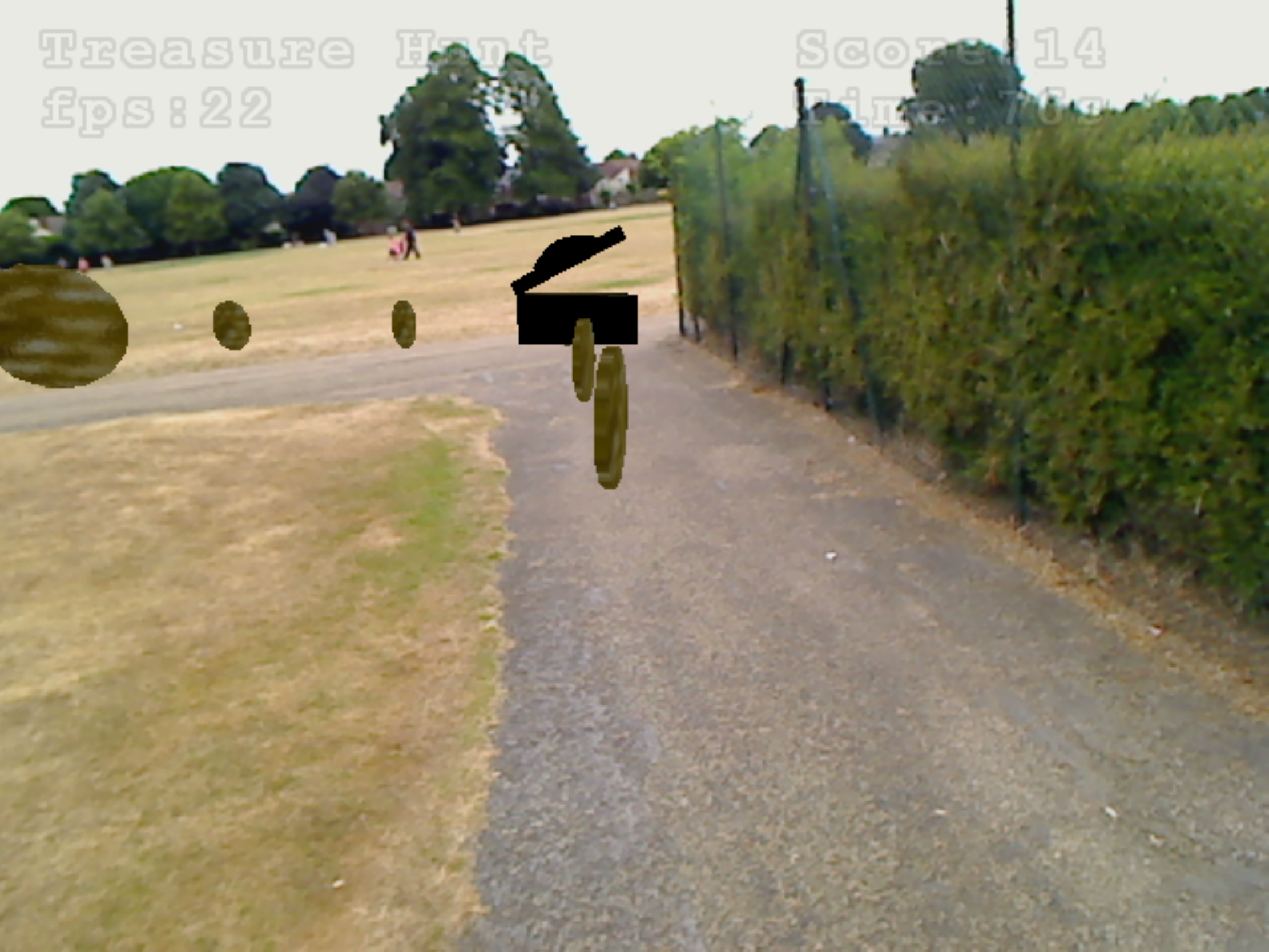}
  \end{center}
  \caption[]{A view from the AR game}
  \label{fig:arGame}
\end{figure}

A view from the AR game is presented in Figure~\ref{fig:arGame}. The game has an interface which displays the score and time passed making the game more challenging and hence interesting. Note that the frame rate of the game is, indeed, very close to video rates (22 frames per second), an indicator of the speed of the filter.

\vfill

\section{Conclusion}
\label{sec:conclusion}
This paper presented a DFA design for motion model selection in GPS--IMU sensor fusion. The results show that multiple-motion model sensor fusion can be achieved by utilising Kalman filter innovation together with a DFA based model selection scheme. It was observed that the use of different motion models can reduce the filter error and prevent divergence. It is clear that choosing the appropriate motion model depending on user's speed improves the accuracy of Kalman filter for tracking applications. 

A sample AR game was used to test the defined approach, and it was observed that the filter is accurate and fast enough to collect all the reward items in the game.

Future work will delve into further analysis of different motion models and a machine learning approach appears to be a promising research direction.

%
%

\ifCLASSOPTIONcaptionsoff
  \newpage
\fi

\bibliographystyle{unsrt} 
\bibliography{References}

%
%

\end{document}